\documentclass[preprint,sort&compress,12pt]{elsarticle}

\usepackage{amssymb}
\usepackage{amsthm}

\begin{document}

\begin{frontmatter}



\title{Composite-particles (Boson, Fermion) Theory of Fractional Quantum Hall Effect}
 

\author[label1]{Shigeji Fujita}
\author{Akira Suzuki\corref{cor2}\fnref{label2}}

\author[label3]{H.C.\,Ho}
\address[label1]{Department of Physics, University at Buffalo, State University of New York,
Buffalo, New York 14260-1500, USA}
\address[label2]{Department~of~Physics, Faculty~of~Science, Tokyo~University~of~Science, 
 Shinjyuku-ku, Tokyo~162-8601, Japan}
\address[label3]{Sincere Learning Centre, Kowloon, Hong Kong, China} 
\date{\today}

\begin{abstract}
A quantum statistical theory is developed for a fractional quantum Hall effects in terms of composite bosons (fermions) each of which contains a conduction electron and an odd (even) number of fluxons.  The cause of the QHE is by assumption the phonon exchange attraction between the conduction electron (``electron", ``hole") and fluxons (quanta of magnetic fluxes).  We postulate that c-fermions with \emph{any} even number of fluxons have an effective charge (magnitude) equal to the electron charge $e$.  The density of c-fermions with $m$ fluxons, $n_\phi^{(m)}$, is connected with the electron density $n_{\mathrm e}$ by $n_\phi^{(m)}=n_{\mathrm e}/m$, which implies a more difficult formation for higher $m$, generating correct values $me^2/h$ for the Hall conductivity $\sigma_{\mathrm H}\equiv j/E_{\mathrm H}$.  
For condensed c-bosons the density of c-bosons-with-$m$ fluxons, $n_\phi^{(m)}$, is connected with the boson density $n_0$ by $n_\phi^{(m)}=n_0/m$.  This yields $\sigma_{\mathrm H}=m\,e^2/h$ for the magnetoconductivity, the value observed of the QHE at filling factor $\nu=1/m$ ($m=$odd numbers).  Laughlin's theory and results about the fractional charge are not borrowed in the present work.
\end{abstract}


\end{frontmatter}


%
\section{Introduction}\label{sec1}
%
In 1983 Laughlin introduced a Laughlin ground-state wave function for a system of $N$ electrons at $\nu=1/3$, and studied the elementary excitations (quasiparticles) over the ground-state\,\cite{1}.  He predicted that the quasiparticle has a \emph{fractional charge}:
\begin{equation}
e^*=e/3\,.\label{1}
\end{equation}
This prediction was later confirmed in the magnetotransport experiments by Clark \emph{et al}.  and others\,\cite{2}.  The 1998 Nobel prize was shared by Tsui, St\"ormer (experimental discovery) and Laughlin (theory) for their contribution to the fractional QHE.

The prevalent theories\,\cite{4} based on the Laughlin wave function in the Schr\"odinger picture deal with the QHE at 0 K and immediately above.  The system ground-state, however, does not carry a current.  To interpret the experimental data it is convenient to introduce composite (c-) particles (bosons, fermions).  The c-boson (fermion), each containing an electron and an odd (even) number of \emph{flux quanta (fluxons)}, were introduced by Zhang \emph{et al}.\,\cite{5} and others (Jain\,\cite{6}) for the description of the fractional QHE (Fermi liquid).  All QHE states with distinctive plateaus in $\rho_{\mathrm H}$ are observed below the critical temperature $T_{\mathrm c}\simeq 0.5$\,K.  The QHE in graphene, a single sheet of graphite, is an exception.  It is desirable to treat the QHE below and above $T_{\mathrm c}$ in a unified manner.  The extreme accuracy (precision $\sim\,10^{-8}$) in which each plateau is observed means that the current density $j$ must be computed exactly without averaging.  In the prevalent theories\,\cite{4} the electron-electron interaction and Pauli's exclusion principle are regarded as the cause for the QHE.  Both are essentially repulsive and cannot account for the fact that the c-particles are bound, that is, they are in negative-energy states.  Besides, the prevalent theories have limitations:
\begin{itemize}
\item The zero temperature limit is taken at the outset.  Then the question why QHE is observed below 0.5\,K in GaAs/AlGaAs cannot be answered.  We better have a theory for all temperatures.
\item The high-field limit is taken at the outset.  The integer QHE are observed for small integer $P$ only.  The question why the QHE is not observed for high $P$ (weak field) cannot be answered.  We better describe the phenomena for all fields.
\item The Hall resistivity $\rho_{\mathrm H}$ value $(Q/P)(h/e^2)$ is obtained in a single stroke.  To obtain $\rho_{\mathrm H}$ we need two separate measurements of the Hall field $E_{\mathrm H}$ and the current density $j$.  We must calculate $(E_{\mathrm H}, j)$ and take the ratio $E_{\mathrm H}/j=\rho_{\mathrm H}$.
\end{itemize}

The main purpose of the present work is to show that the known fractional QHE can be described within the frame-work of c-particles model without using Laughlin's theory and results about the fractional charge.
%
\section{The Hamiltonian}\label{sec2}
%
Fujita and Okamura developed a quantum statistical theory of the QHE\,\cite{6}.  We follow this theory.  See this reference for more details.  

There is a remarkable similarity between the QHE \emph{and} the {High-Temperature Superconductivity} (HTSC), both occurring in two-dimensional (2D) systems as pointed out by Laughlin\,\cite{7}.  The major superconducting properties observed in the HTSC are (a) zero resistance, (b) a sharp phase change, (c) an energy gap below $T_{\rm c}$, (d) flux quantization, (e) Meissner effect, and (f) Josephson effects.  All these have been observed in GaAs/AlGaAs.  We regard the \emph{phonon exchange attraction} as the causes of both QHE and superconductivity.  Starting with a reasonable Hamiltonian, we calculate everything using quantum statistical mechanics. 

The countability concept of the fluxons, known as the \emph{flux quantization}:
\begin{equation}
B=\frac{N_\phi}{A}\frac{h}{e}\equiv n_\phi\frac{h}{e}\,,\label{2}
\end{equation}
where $A=$ sample area, $N_\phi=$ fluxon number (integer) and $h=$ Planck constant, is originally due to Onsager\,\cite{8}.  The magnetic (electric) field is an axial (polar) vector and the associated fluxon (photon) is a half-spin fermion (full-spin boson).  The magnetic (electric) flux line cannot (can) terminate at a sink, which supports the fermionic (bosonic) nature of the associated fluxon (photon).  No half-spin fermion can annihilate by itself because of angular momentum conservation.  The electron spin originates in the relativistic quantum equation (Dirac's theory of electron)\,\cite{9}.  The discrete (two) quantum numbers $(\sigma_z=\pm1)$ cannot change in the continuous limit, and hence the spin must be conserved.  The countability and statistics of the fluxon are fundamental particle properties.  We postulate that \emph{the fluxon is a half-spin fermion with zero mass and zero charge}.

We assume that the magnetic field $\mathbf B$ is applied perpendicular to the interface.  The 2D Landau level energy,
\begin{equation}
\varepsilon=\hbar\omega_{\mathrm c}\left(N_{\mathrm L}+\frac12\right)\,,\quad\omega_{\mathrm c}\equiv eB/m^*\,,\label{3}
\end{equation}
with the states $(N_{\mathrm L}, k_y)$, $N_{\mathrm L}=0,\,1,\,2,\,\cdots$, have a great degeneracy.  The cyclotron frequency $\omega_{\mathrm c}$ contains the electron effective mass $m^*$.  The Center-of-Mass (CM) of \emph{any} c-particle moves as a fermion (boson).  The eigenvalues of the CM momentum are limited to 0 or 1 (unlimited) if it contains an odd (even) number of elementary fermions.  This rule is known as the \emph{Ehrenfest-Oppenheimer-Bethe's} (EOB's) \emph{rule}\,\cite{10}. Hence the CM motion of the composite containing an electron and $Q$ fluxons is bosonic (fermonic) if $Q$ is odd (even).  The system of c-bosons condenses below the critical temperature $T_{\mathrm c}$ and exhibits a superconducting state while the system of c-fermions shows a Fermi liquid behavior.

A longitudinal phonon, acoustic or optical, generates a density wave, which affects the electron (fluxon) motion through the charge displacement (current).  The exchange of a phonon between electrons and fluxons generates an \emph{attractive} transition.

Bardeen, Cooper and Schrieffer (BCS)\,\cite{11} assumed the existence of Cooper pairs\,\cite{12} in a superconductor, and wrote down a Hamiltonian containing the ``electron" and ``hole" kinetic energies and the pairing interaction Hamiltonian with the phonon variables eliminated.  We start with a BCS-like Hamiltonian ${\mathcal H}$ for the QHE\,\cite{6}:
\begin{eqnarray}
{\mathcal H}\!&=&\!{\sum_{\mathbf k}}^{\prime}\sum_{s}\varepsilon_{\mathbf k}^{(1)}n_{{\mathbf k}s}^{(1)}
+{\sum_{\mathbf k}}^{\prime}\sum_{s}\varepsilon_{\mathbf k}^{(2)}n_{{\mathbf k}s}^{(2)}+{\sum_{\mathbf k}}^{\prime}\sum_{s}\varepsilon_{\mathbf k}^{(3)}n_{{\mathbf k}s}^{(3)}-{\sum_{\mathbf q}}^{\prime}{\sum_{\mathbf k}}^{\prime}\nonumber\\
&&{\sum_{{\mathbf k}^{\prime}}}^{\prime}
\sum_{s}v_{0}\left[ B_{{\mathbf k}^{\prime}{\mathbf q}\,s}^{(1)\dagger}B_{{\mathbf k}{\mathbf q}\,s}^{(1)}+B_{{\mathbf k}^{\prime}{\mathbf q}\,s}^{(1)\dagger}B_{{\mathbf k}{\mathbf q}\,s}^{(2)\dagger}+B_{{\mathbf k}^{\prime}{\mathbf q}\,s}^{(2)}B_{{\mathbf k}{\mathbf q}\,s}^{(1)}+B_{{\mathbf k}^{\prime}{\mathbf q}\,s}^{(2)}B_{{\mathbf k}{\mathbf q}\,s}^{(2)\dagger} \right],\quad\label{4}
\end{eqnarray}
where $n_{{\mathbf k}s}^{(j)}=c_{{\mathbf k}s}^{(j)\dagger}c_{{\mathbf k}s}^{(j)}$ is the number operator for the ``electron'' (1) [``hole'' (2), fluxon (3)] at momentum ${\mathbf k}$ and spin $s$ with the energy $\varepsilon_{{\mathbf k},s}^{(j)}$, with annihilation (creation) operators $c$ $(c^\dagger)$ satisfying the Fermi anti-commutation rules:
\begin{equation}
\{c_{{\mathbf k}s}^{(i)},\,c_{{\mathbf k}^{\prime}s^{\prime}}^{(j)\dagger}\}
\equiv  c_{{\mathbf k}s}^{(i)}c_{{\mathbf k}^{\prime}s^{\prime}}^{(j)\dagger} + c_{{\mathbf k}^{\prime}s^{\prime}}^{(j)\dagger}c_{{\mathbf k}s}^{(i)}=\delta_{{\mathbf k},{\mathbf k}^{\prime}}\delta_{s,s^{\prime}}\delta_{i,j}\,,\quad\{c_{{\mathbf k}s}^{(i)}, c_{{\mathbf k}^{\prime}s^{\prime}}^{(j)}\}=0\,.\label{5}
\end{equation}
The fluxon number operator $n_{{\mathbf k}s}^{(3)}$ is represented by $a_{{\mathbf k}s}^\dagger a_{{\mathbf k}s}$ with $a$ $(a^\dagger)$ satisfying  the anti-commutation rules:
\begin{equation}
\{a_{{\mathbf k}s},\,a_{{\mathbf k}^{\prime}s^{\prime}}^\dagger\} = \delta_{{\mathbf k},{\mathbf k}^{\prime}}\delta_{s,s^{\prime}}\,,\quad\{a_{{\mathbf k}s},\,a_{{\mathbf k}^{\prime}s^{\prime}}\}=0\,.\label{6}
\end{equation}

The phonon exchange can create electron-fluxon composites, bosonic or fermionic, depending on the number of fluxons.  The center-of-mass of any composite moves as a fermion\,(boson) if it contains an odd\,(even) numbers of elementary fermions.  We call the conduction-electron composite with an odd (even) number of fluxons c-boson\,(c-fermion).  The electron\,(hole)-type c-particles carry negative (positive) charge.  We expect that electron\,(hole)-type c-bosons are generated by the phonon-exchange attraction.  The pair operators  $B$ are defined by
 \begin{equation}
 B_{{\mathbf{kq}},s}^{(1)\dagger}\equiv c_{{\mathbf k}+{\mathbf q}/2,s}^{(1)\dagger}a_{-{\mathbf k}+{\mathbf q}/2,-s}^{\dagger}\,,\quad B_{{\mathbf{kq}},s}^{(2)}\equiv a_{-{\mathbf k}+{\mathbf q}/2,-s}c_{{\mathbf k}+{\mathbf q}/2,s}^{(2)}\,.\label{7}
 \end{equation}
The prime on the summation in Eq.\,(\ref{4}) means the restriction: $0<\varepsilon_{{\mathbf k}s}^{(j)}<\hbar\omega_{\rm D}$, $\omega_{\rm D}=$ Debye frequency.    
The pairing interaction terms in Eq.\,(\ref{4}) conserve the charge.  The term $-v_{0}B_{{\mathbf k}^{\prime}{\mathbf q}\,s}^{(1)\dagger}B_{{\mathbf k}{\mathbf q}\,s}^{(1)}$, where $v_{0}\equiv \vert V_{{\mathbf q}}V_{{\mathbf q}}^{\prime}\vert\,(\hbar\omega_{0}A)^{-1}$, $A=$ sample area, is the pairing strength, generates a transition in electron-type c-fermion states.  Similarly, the exchange of a phonon generates a transition between hole-type c-fermion states, represented by $-v_{0}B_{{\mathbf k}^{\prime}{\mathbf q}\,s}^{(2)\dagger}B_{{\mathbf k}{\mathbf q}\,s}^{(2)\dagger}$. 
 The phonon exchange can also pair-create\,(pair-annihilate) electron\,(hole)-type c-boson pairs, and the effects of these processes are represented by $-v_{0}B_{{\mathbf k}^{\prime}{\mathbf q}\,s}^{(1)\dagger}B_{{\mathbf k}{\mathbf q}\,s}^{(2)\dagger}$ $\left(-v_{0}B_{{\mathbf k}{\mathbf q}\,s}^{(1)}B_{{\mathbf k}{\mathbf q}\,s}^{(2)}\right)$.  
 
The Cooper pair is formed from two ``electrons" (or ``holes").  Likewise the c-bosons may be formed by the phonon-exchange attraction from c-fermions and fluxons.  If the density of the c-bosons is high enough, then the c-bosons will be condensed and exhibit a superconductivity. 

To treat  superconductivity we modify the pair operators in Eq.\,(\ref{7}) as  
 \begin{equation}
 B_{{\mathbf{kq}},s}^{(1)\dagger}\equiv c_{{\mathbf k}+{\mathbf q}/2,s}^{(1)\dagger}c_{-{\mathbf k}+{\mathbf q}/2,-s}^{(1)\dagger}\,,\quad B_{{\mathbf{kq}},s}^{(2)}\equiv c_{-{\mathbf k}+{\mathbf q}/2,-s}^{(2)}c_{{\mathbf k}+{\mathbf q}/2,s}^{(2)}\,.\label{8}
\end{equation}
Then, the pairing interaction terms in Eq.\,(4) are formally identical with those in the generalized BCS Hamiltonian\,\cite{13}.

We first consider integer QHE.  We choose a conduction electron and a fluxon  for the pair.  The c-bosons, having the linear dispersion relation, can move in all directions in the plane with the constant speed $(2/\pi)v_{\mathrm F}^{(j)}$.  The supercurrent is generated by $\mp$\,c-bosons monochromatically condensed, running along the sample length.  The supercurrent density (magnitude) $j$, calculated by the rule:  $j=(\hbox{carrier charge}\,e^*)\times(\hbox{carrier density}\,n_0)\times(\hbox{drift velocity}\,v_{\mathrm d})$, is given by
\begin{equation}
j\equiv e^{*}n_{0}v_{\mathrm d}=e^*n_0\,\frac{2}{\pi}\left\vert v_{\mathrm F}^{(1)} - v_{\mathrm F}^{(2)}\right\vert\,.\label{9}
\end{equation}
The induced Hall field (magnitude) $E_{\mathrm H}$ equals $v_{\mathrm d}B$.  The magnetic flux is quantized:
\begin{equation}
B = n_\phi\Phi_0\,, \label{10}
\end{equation}
where $n_{\phi}\equiv N_\phi/A$ is the fluxon density and $\Phi_0\,(\equiv\,e/h)$ is the magnetic flux quantum.  
Hence we obtain
\begin{equation}
\rho_{\mathrm H}\equiv\frac{E_{\mathrm H}}{j}=\frac{v_{\mathrm d}B}{e^*n_0v_{\mathrm d}}
=\frac{1}{e^{*}n_{0}}n_{\phi}\left(\frac{h}{e}\right)\,.
\label{11}
\end{equation}
We assume that \emph{the c-fermion containing an electron and an even number of fluxons has a charge magnitude} $e$.  For the integer QHE, $e^{*}=e$, $n_{\phi}=n_{0}$ for the carriers, thus we obtain $\rho_{\rm H}=h/e^2$, the correct plateau value observed for the  principal QHE at $\nu=1$.

The supercurrent generated by equal numbers of $\mp$\,c-bosons condensed monochromatically is neutral.  This is reflected in our calculations in Eq.\,(\ref{9}).  The supercondensate whose motion generates the supercurrent must be neutral.  If it has a charge, it would be accelerated indefinitely by the external field because the impurities and phonons cannot stop the supercurrent to grow.  That is, the circuit containing a superconducting sample and a battery must be burnt out if the supercondensate is not neutral. 
In the calculation of $\rho_{\mathrm H}$ in Eq.\,(\ref{11}), we used the \emph{unaveraged} drift velocity difference $(2/\pi)\vert v_{\mathrm F}^{(1)}-v_{\mathrm F}^{(2)}\vert$, which is significant.  \emph{Only} the unaveraged drift velocity cancels out $v_{\mathrm d}$ exactly from numerator/denominator, leading to an exceedingly accurate plateau value.  Thus we derived the precise plateau value $h/e^{2}$ in experiment for the  principal QHE

We now extend our theory to include elementary fermions (electron, fluxon) as members of the c-fermion set.  We can then treat the superconductivity and the QHE in a unified manner.  The c-boson containing a pair of one electron and one fluxon can be used to describe the room temperature QHE in graphene\,\cite{14}.

Important pairings and effects are listed below.
\begin{itemize}
\item[(a)] a pair of conduction electrons, superconductivity
\item[(b)] fluxon and c-fermions, QHE
\item[(c)] a pair of like-charge conduction electrons with two fluxons, QHE in graphene. 
\end{itemize}
%
\section{Fractional Quantum Hall Effect}\label{sec3}
%
We postulate that \emph{any} c-fermion has the effective charge $e^*$ equal to the electron charge (magnitude) $e$:
\begin{equation}
e^*=e\quad\hbox{for \emph{any} c-fermion}.\label{12}
\end{equation}
Let us consider a c-fermion containing an electron and two fluxons.  We shall show that the c-fermion density $n_\phi^{(2)}$ is connected with the electron density $n_{\mathrm e}$ by
\begin{equation}
n_\phi^{(2)}=n_{\mathrm e}/2\,.\label{13}
\end{equation}

If the c-fermions run in the $E$-field direction with the drift velocity $v_{\mathrm d}$, the current density $j$ is given by
\begin{equation}
j=e^*n_{\mathrm e}v_{\mathrm d}\,.\label{14}
\end{equation}
The Hall field $E_{\mathrm H}$ is 
\begin{equation}
E_{\mathrm H}=v_{\mathrm d}B\,.\label{15}
\end{equation}
The Hall conductivity $\sigma_{\mathrm H}$ is defined and calculated as follows.
\begin{eqnarray}
\sigma_{\mathrm H}&\equiv&\frac{j}{E_{\mathrm H}}=\frac{e^*n_0v_{\mathrm d}}{v_{\mathrm d}B}
=\frac{e^*n_{\mathrm e}}{n_\phi^{(2)}(h/e)}\nonumber\\
&=&e^*e\frac{n_{\mathrm e}}{n_\phi^{(2)}(h/e)} =\frac{e^*e}{h}\frac{n_{\mathrm e}}{n_\phi^{(2)}}=\frac{2e^2}{h}\,,\label{16}
\end{eqnarray}
which is the correct value of $\sigma_{\mathrm H}$ at $\nu=1/2$.  The last two members of  Eqs .\,(\ref{16}) signifies Eq.\,(\ref{13}).  In fact if we assume Eqs.~ (\ref{12}) and (\ref{13}), then we obtain Equations (\ref{16}).

Similarly for the case of c-fermions with four fluxons we obtain $\sigma_{\mathrm H}=4e^2/h$.

Extending Eq.\,(13) to a general case, we obtain
\begin{equation}
n_\phi^{(m)}=n_{\mathrm e}/m\,,\label{17}
\end{equation}
where $m$ is an odd number.  The density $n_\phi^{(m)}$ is proportional to the magnetic field $B$.  Equation (\ref{17}) is valid for a large $m$.  As the magnetic field is raised, the separation between the LL  becomes greater, and the higher-$m$ c-fermion is more difficult to form energetically.  This condition is unlikely to depend on the statistics of the c-particles.  Thus Eq.\,(\ref{17}) should be valid for all integers, odd or even.

We take the case of $m=3$.  The c-boson containing an electron and three fluxons can be formed from a c-fermion with two fluxons and a fluxon by the phonon exchange attraction.  If the c-bosons are Bose-condensed, the supercurrent density $j$ is given by Eq.\,(\ref{9}).  Hence we obtain
\begin{eqnarray}
\rho_{\mathrm H}&\equiv&\frac{E_{\mathrm H}}{j}=\frac{v_{\mathrm d}B}{e^*n_0v_{\mathrm d}}=\frac{n_\phi^{(3)}}{e^*n_0}\left(\frac he\right)\nonumber\\
&=&\frac13\frac{h}{e^2}\,,\label{18}
\end{eqnarray}
where we used $e^*=e$ from Eq.\,(\ref{12}), and $n_\phi^{(3)}/n_0=1/3$, a bosonic extension of Eq.\,(\ref{17}).

The principal fractional QHE occurs at $\nu=1/3$, where the Hall resistivity value is $h/3e^2$.  A set of weaker QHE occur on the weaker field side at
\begin{equation}
\nu=1/3,\,2/3,\,\cdots\,.\label{19}
\end{equation}
The QHE behavior at $\nu=P/Q$ for any $Q$ is similar.  We illustrate it by taking integer QHE with $\nu=P=1,\,2,\,\cdots$.  The field magnitude becomes smaller with increasing $P$.  The LL degeneracy is proportional to $B$, and hence $P$ LL's must be considered.  First consider the case $P=2$.  Without the phonon-exchange attraction the electrons occupy the lowest two LL's with spin.  The electrons at each level form fundamental (f)\,c-bosons.   
%
\begin{figure}[htbp]
\begin{center}
\includegraphics[scale=1.0]{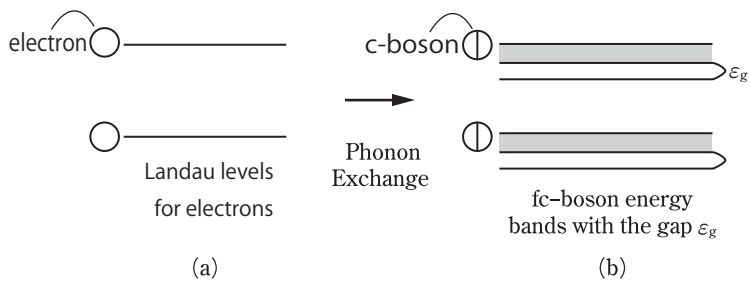}
\caption{The electrons which fill up the lowest two LL's, form 
the QH state at $\nu=2$ after the phonon-exchange attraction 
and the BEC of the c-bosons.}
\label{fig1}
\end{center}
\end{figure}
%
In the superconducting state the supercondensate occupy the monochromatically condensed state, which is separated by the superconducting gap $\varepsilon_{\mathrm g}$ from the continuum states (band) as shown in the right-hand figure in Fig.\,1.

The c-boson density $n_0$ at each LL is one-half the density at $\nu=1$, which is equal to the 
electron density $n_{\mathrm e}$ fixed for the sample.  Extending the theory to a general integer, we have
\begin{equation}
n_0=n_{\mathrm e}/P\,.\label{20}
\end{equation}
This means that both $T_{\mathrm c}\,(\propto n_0^{1/2})$ and $\varepsilon_{\mathrm g}$ are smaller, making the plateau width (a measure of $\varepsilon_{\mathrm g}$) smaller in agreement with experiments.  The c-bosons have lower energies than the conduction electrons.  Hence at the extreme low temperatures the supercurrent due to the condensed c-bosons dominates the normal current due to the conduction electrons and non-condensed c-bosons, giving rise to the dip in $\rho$.
%
\section{Summary and Discussion}\label{sec4}
%
In summary we have achieved our goal of treating fractional QHE in terms of c-particles without using Laughlin's results in terms of fractional charges carried by the quasiparticles.  We found that the principal fractional QHE occurs at $\nu=1/Q$, $Q$ (odd integers) with the Hall conductivity $\sigma_{\mathrm H}=Qe^2/h$, where the density of c-bosons with $Q$ fluxons, $n_\phi^{(Q)}$, is connected with the density of condensed c-bosons with $Q$ fluxons, $n_0$, by $n_\phi^{(Q)}=n_0/Q$. 

Other significant findings are:
\begin{itemize}
\item A set of weaker fractional QHE occur on the smaller field side at $\nu=P/Q$, with the Hall conductivity $\sigma_{\mathrm H}=(Q/P)\,e^2/h$.  The plateau widths (a measure of $\varepsilon_{\mathrm g}$) are smaller are since the condensed c-boson density $n_0$ is smaller.  Both the critical temperature $T_{\mathrm c}\,\,(\propto n_0^{1/2})$ and the superconducting energy gap $\varepsilon_{\mathrm g}$ are smaller.
\item The fractional fermonic principal QHE with a finite conductivity and the Hall conductivity $\sigma_{\mathrm H}$ equal to $Q\,(e^2/h)$, occur at $\nu=1/Q$, $Q$ (even integers), where the density of c-fermions with $Q$ fluxons, $n_\phi^{(Q)}$, is connected with the density of conduction electrons, $n_{\mathrm e}$, by $n_\phi^{(Q)}=n_{\mathrm e}/Q$.  All of the QHE points fall on the straight line representing the classical Hall effect if the $\sigma_{\mathrm H}$ is plotted as a function of $\nu=1/Q$.
\item The higher in $m$ c-particle is more difficult to form energetically since the LL separation is greater.
\end{itemize}
In the previous work\,\cite{14} we showed that 
\begin{itemize}
\item the horizontal stretch of $\sigma_{\mathrm H}$ accompanied by the zero resistivity at the mid-point of the stretch, the signature of the QHE, arises from the superconducting energy gap $\varepsilon_{\mathrm g}$.
\item The cause of the QHE is the phonon exchange attraction between c-fermion and fluxons.  This allows us to develop a unified theory of superconductivity and QHE.  
\item The Hall conductivity $\sigma_{\mathrm H}$ ($\equiv$ the current density $j$/the Hall field $E_{\mathrm H}$) can be calculated exactly with the assumption of the BEC of the c-bosons.
\end{itemize}
Our quantum statistical theory is a finite temperature theory.  The room-temperature QHE in graphene is an important  topic, which is discussed separately. See Ref.\,\cite{14}.
%
%

%
\end{document}